\documentclass[12pt]{iopart}
\usepackage{epsfig}  
\usepackage{caption}
\begin{document}

\def\NZ{{\bar N_0}}
\def\NJ{N_{\J}}
\def\J{J/\psi}
\newcommand{\Npart}{N_{part}}
\newcommand{\Nc}{N_c}
\newcommand{\Nch}{N_{ch}}
\newcommand{\Ncbar}{N_{\bar c}}
\newcommand{\ccbar}{c \bar c}
\newcommand{\Nccbar}{N_{\ccbar}}
\newcommand{\kts}{\langle{k_T}^2\rangle}
\newcommand{\pts}{\langle{p_T}^2\rangle}
\newcommand{\epsz}{\epsilon_0}
\newcommand{\sigF}{\sigma_F}
\newcommand{\sigD}{\sigma_D}
\def\pt{$p_{T}~$}
\def\RAA{$R_{AA}$}
\newcommand{\gev}{\mathrm{GeV}}
\title{In-medium formation of quarkonium} 

\author{R. L. Thews  
}

\address{Department of Physics, University of Arizona,
Tucson, AZ 85721 USA}



\begin{abstract}
We confront preliminary RHIC data on $\J$ production in 
nuclear interactions with expectations which follow in 
scenarios involving charm quark recombination
in a region of color deconfinement.  The focus is on 
transverse momentum and rapidity spectra of the $\J$, 
which carry a memory of the spectra of the 
charm quarks.  In such a scenario, one predicts that 
both spectra
will be narrower than those
expected without recombination.  Preliminary results 
for the transverse momentum spectra point toward a 
preference for the recombination picture, while the
rapidity spectra do not exhibit any narrowing within present
large uncertainties.  We present new calculations in the recombination
model for the centrality behavior of these signals, which map out 
the necessary experimental precision required for a
definitive test.

\end{abstract}




\section{Introduction}

Recent PHENIX results on $\J$ production in pp, dA, and AA collisions
at RHIC provide a first opportunity to test for the presence of
in-medium formation, which
is expected to become significant when multiple pairs
of heavy quarks are present in a single AA collision \cite{jpsiform}.
Preliminary results \cite{PereiraDaCosta:2005xz}
for $R_{AA}$ as a function of centrality
show less suppression than expected in several model calculations, but 
can be reproduced by revised models in which the initially-produced 
$\J$ are almost completely suppressed, and are then augmented 
by an additional component which can be 
attributed to in-medium formation. 
However the precise magnitude of in-medium formation is in general sensitive
to several model parameters which are not at present well-constrained 
from independent measurements.
An alternative scenario asserts that there is no dissociation of primary 
$\J$ even at RHIC energy \cite{Karsch:2005nk}. Thus one must await results at
LHC energy to test the sequential suppression scenario using only the
magnitude of $\J$ production.

Our recent  work in this area concentrated on finding a signature
for in-medium $\J$ formation which is
independent of the detailed dynamics and magnitude of
the formation.  We found that the {\it normalized} \pt spectrum of the
formed $\J$ may provide such a signature \cite{Thews:2005vj},
because the spectrum of $\J$ formed in-medium is closely
correlated with the spectrum of $\ccbar$ pairs from which they
are formed.  The difference between $\J$ formed in-medium 
and those which are produced in the initial stages is 
due to the combined effect of (a) different properties of the 
``off-diagonal"
pairs compared with the ``diagonal pairs", and (b) inclusion of
a formation probability which characterizes the in-medium process.
The virtue of this approach is that we do not require a knowledge of
the absolute magnitude of in-medium formation in order to isolate
the signatures.

In the next section, we review the properties of a generic kinetic model 
for in-medium recombination and
the resulting formalism to calculate normalized spectra. 
We calculate the initial charm quark spectra using NLO pQCD amplitudes, 
adjusted for
initial state broadening in AA collisions.  This is appropriate in
the case where the interaction between charm quarks and the medium is 
negligible.  
We find that the resulting $\J$ \pt spectra are quite
distinct from that of the initial state diagonal pair sample.
The following section exhibits the model calculations of normalized
spectra. 
More recent work on the centrality behavior of these spectra is 
presented in the next section, along with a comparison with 
preliminary PHENIX measurements.
A brief summary completes this presentation.

\section{Kinetic model of recombination}

We calculate the net number of $\J$ produced in a color-deconfined
medium due to the competing reactions of (a) in-medium formation 
involving recombination
of $c$ and $\bar{c}$ and (b) dissociation of $\J$ induced by interactions
with the medium.  The simplest dissociation reaction
utilizes absorption of single deconfined gluons in the medium to ionize the
color singlet $\J$, $g + \J \rightarrow c + \bar{c}$,
resulting in a $\ccbar$ pair in a color octet state. 
The inverse of this process then serves
as the corresponding formation reaction, in which a $\ccbar$ pair in
a color octet state emits a color octet gluon and falls into the
color singlet $\J$ bound state. Of course, any dissociation mechanism
will have a corresponding in-medium formation mechanism involving the
time-reversed reaction, but the relative magnitudes will depend on
the momentum distribution of the initial state participants.

One can then follow the time evolution
of charm quark and charmonium numbers according
to a Boltzmann equation in which the formation and dissociation
reactions compete.

\begin{equation}\label{eqkin}
\frac{d\NJ}{dt}=
  \lambda_{\mathrm{F}} N_c\, N_{\bar c }[V(t)]^{-1} -
    \lambda_{\mathrm{D}} \NJ\, \rho_g\,,
\label{formationintegral}
\end{equation}
with $\rho_g$ the number density of gluons in the medium.
The reactivity $\lambda$ is the product of the reaction
cross section and initial relative velocity
$\langle \sigma v_{\mathrm{rel}} \rangle$
averaged over the momentum distribution of the initial
participants, i.e. $c$ and $\bar c$ for $\lambda_F$ and
$\J$ and $g$ for $\lambda_D$.
The gluon density is determined by the equilibrium value in the
medium at each temperature, and the volume must be modeled according to the
expansion and cooling profiles of the heavy ion interaction region.

Eq. \ref{formationintegral} has an analytic solution in the case where the total number
of formed $\J$ is much smaller than the initial number of $\Nccbar$.

\begin{equation}
\NJ(t_f) = \epsilon(t_f) [\NJ(t_0) +
\Nccbar^2 \int_{t_0}^{t_f}
{\lambda_{\mathrm{F}}\, [V(t)\, \epsilon(t)]^{-1}\, dt}],
\label{eqbeta}
\end{equation}
where $t_0$ and $t_f$ define the lifetime of the deconfined region.
Note that the function $\epsilon(t_f) = 
e^{-\int_{t_0}^{t_f}{\lambda_{\mathrm{D}}\, \rho_g\,
dt}}$
would be the suppression factor in this scenario if the
formation mechanism were neglected.

One sees that the second term, quadratic in $\Nccbar$, is precisely the
total number of recombinations which occur in the deconfinement volume,
modified by the factor $\epsilon(t_f)/\epsilon(t)$, which is just the
suppression factor for $\J$ formed between times $t$ and $t_f$.

It is obvious that a prediction of total initial plus in-medium
formation population depends on many parameters, not all of which are
constrained by independent information.  Certainly
the number of $\ccbar$ pairs in the region of color deconfinement sets
an important overall scale and centrality dependence.  There are measurements
by STAR \cite{Adams:2004fc} and PHENIX \cite{Adler:2004ta} of these quantities, 
which at present have uncertainties in the factor of 2 range.  At least
as significant is the variation of $\lambda_{\mathrm{F}}$ for 
different charm quark momentum distributions.  We find that this 
quantity decreases by a factor of approximately five when the
distribution changes between thermal equilibrium and that predicted
by initial production as calculated in pQCD.
In addition,
one needs to model the geometric properties of the region of color
deconfinement, including the centrality dependence and accounting for
the expansion profile. An initial temperature $T_0$ controls the
dissociation rate, and also the time evolution of the geometry. Another
parameter (x) specifies the initial number of $\J$ produced in normal
nucleon-nucleon interactions as a fraction of $\Nccbar$. 

\begin{figure}[htb]
\begin{minipage}[t]{80mm}
\epsfig{clip=,width=7.8cm,figure=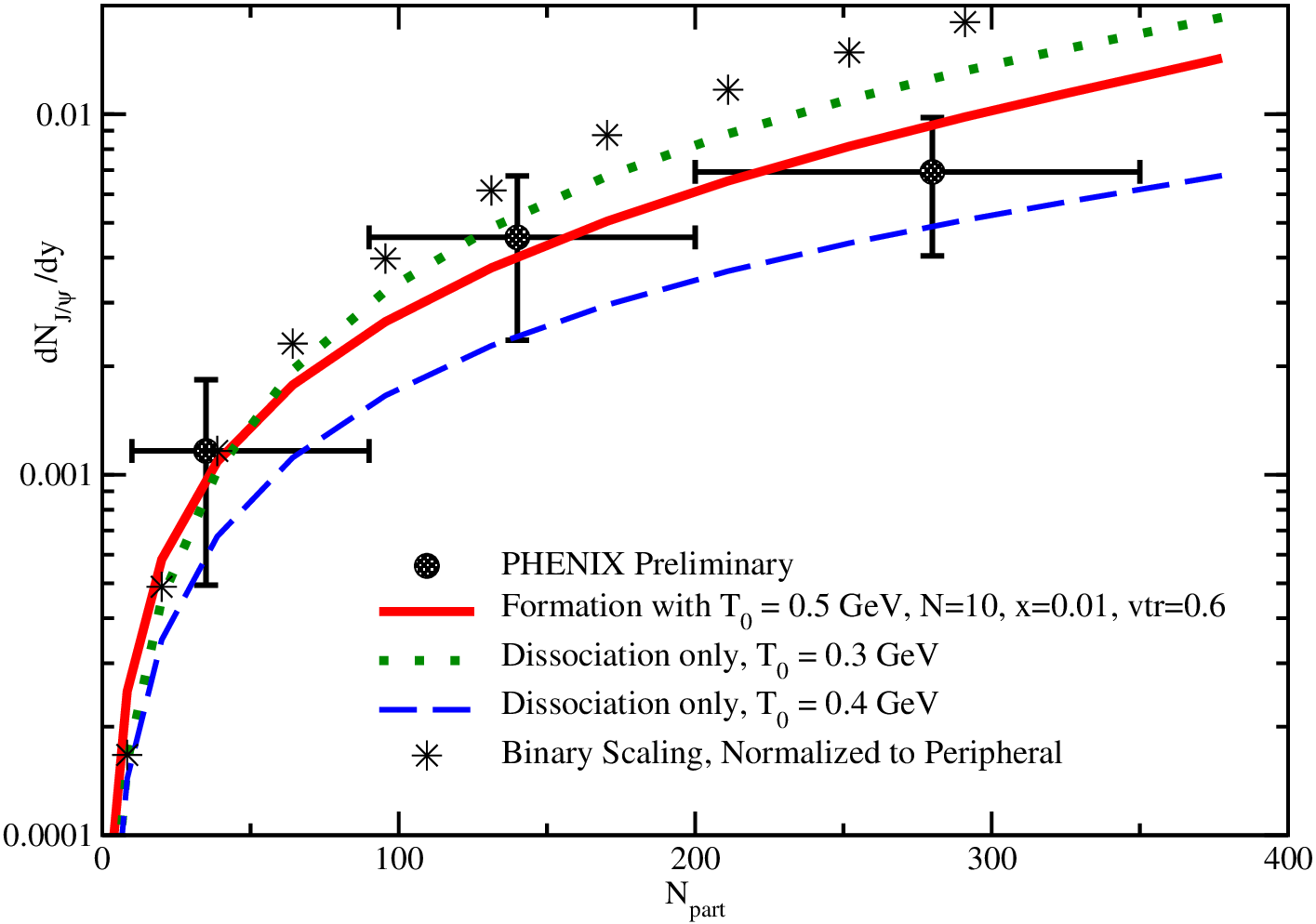}
\caption{\small Centrality dependence of initial data.}
\label{InitialPHENIXdata}
\end{minipage}
\hspace{\fill}
\begin{minipage}[t]{80mm}
\epsfig{clip=,width=7.8cm,figure=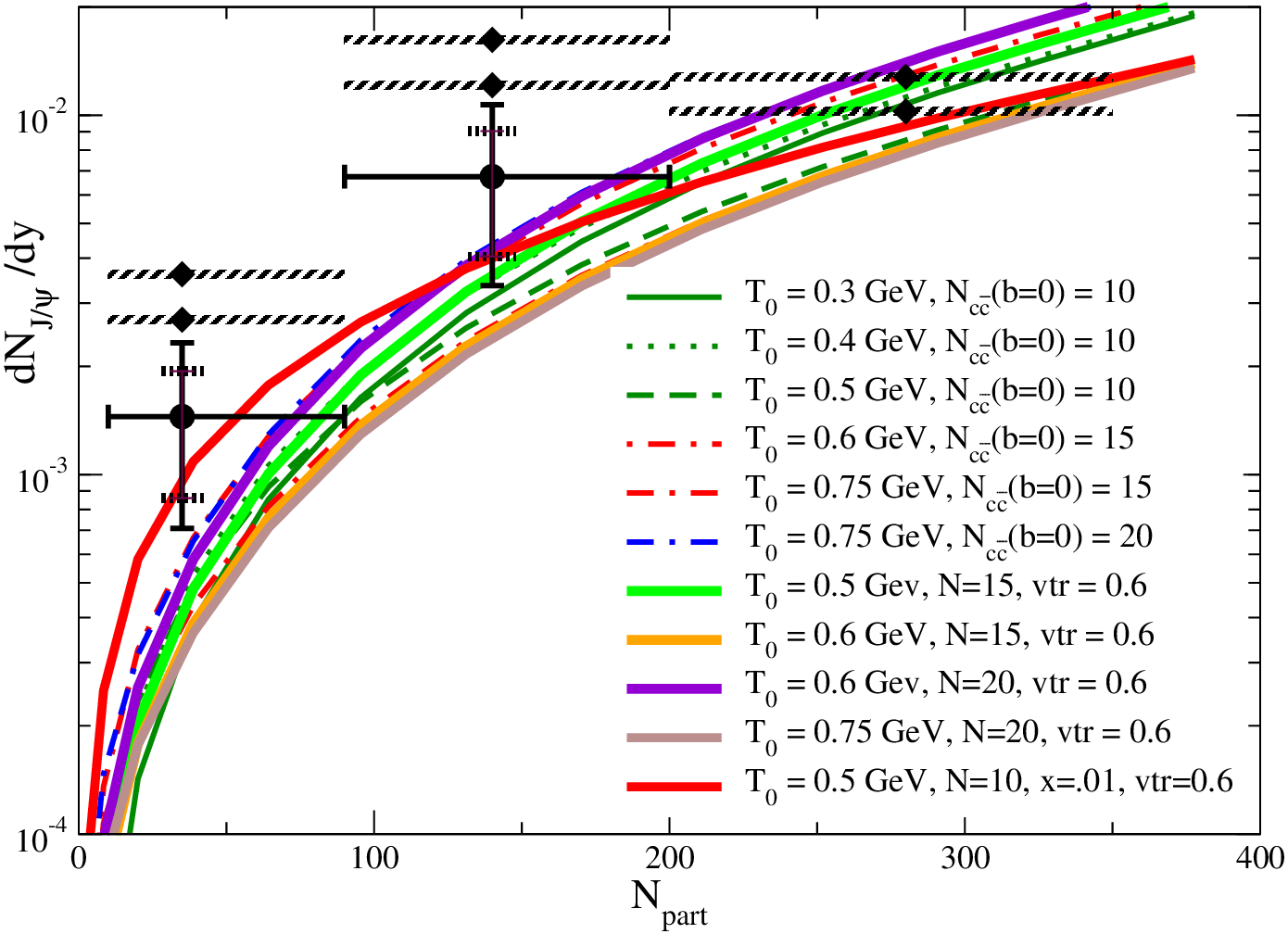}
\caption{\small Exploration of formation model parameter space allowed by
initial data.}
\label{parameterspaceconstraint}
\end{minipage}
\end{figure}

\begin{figure}[htb]
\begin{minipage}[t]{80mm}
\epsfig{clip=,width=7.8cm,figure=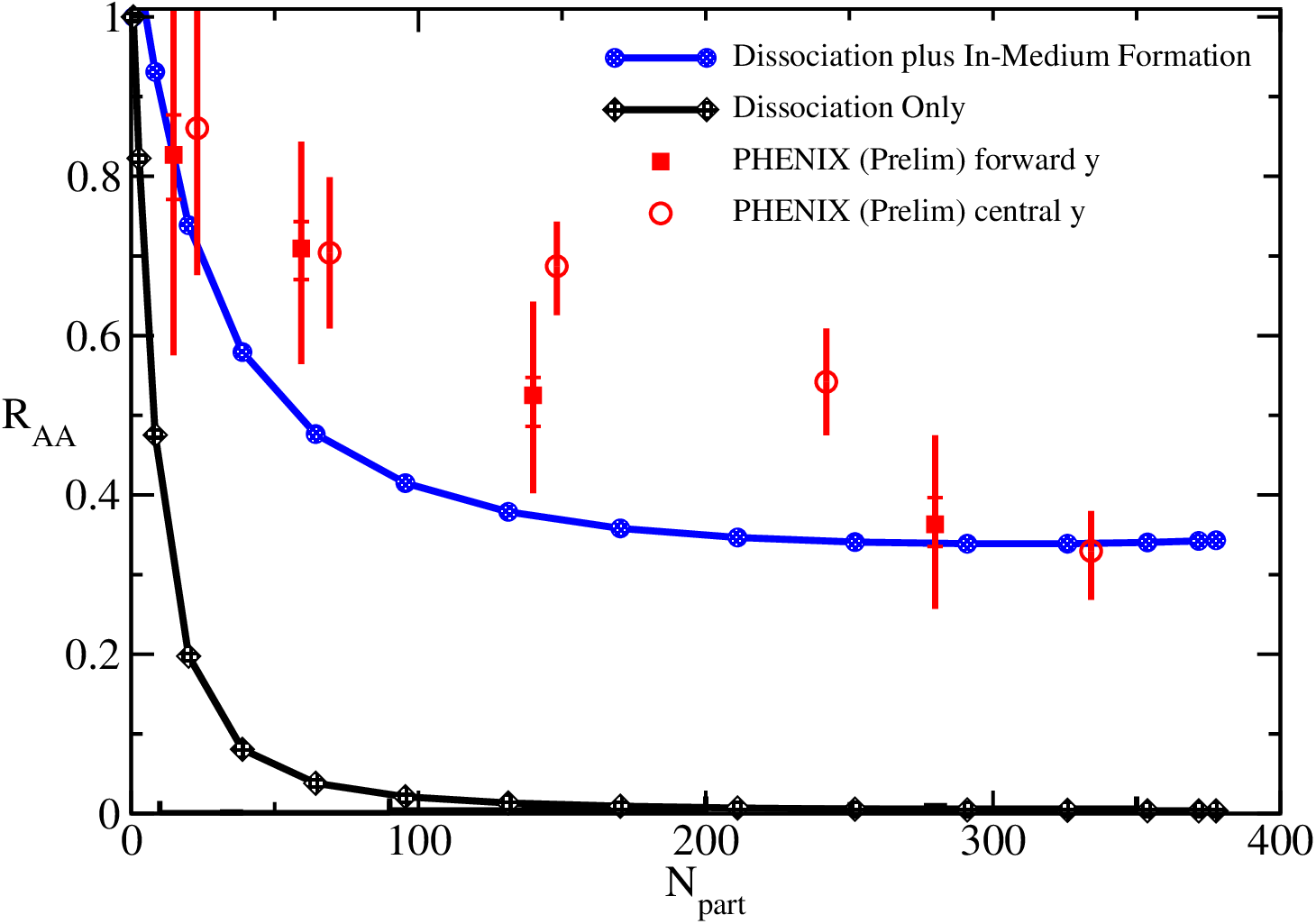}
\caption{\small Centrality dependence of \\kinetic model 
predictions for $R_{AA}$ of $\J$.}
\label{kineticRAApredictions}
\end{minipage}
\hspace{\fill}
\begin{minipage}[t]{80mm}
\epsfig{clip=,width=7.8cm,figure=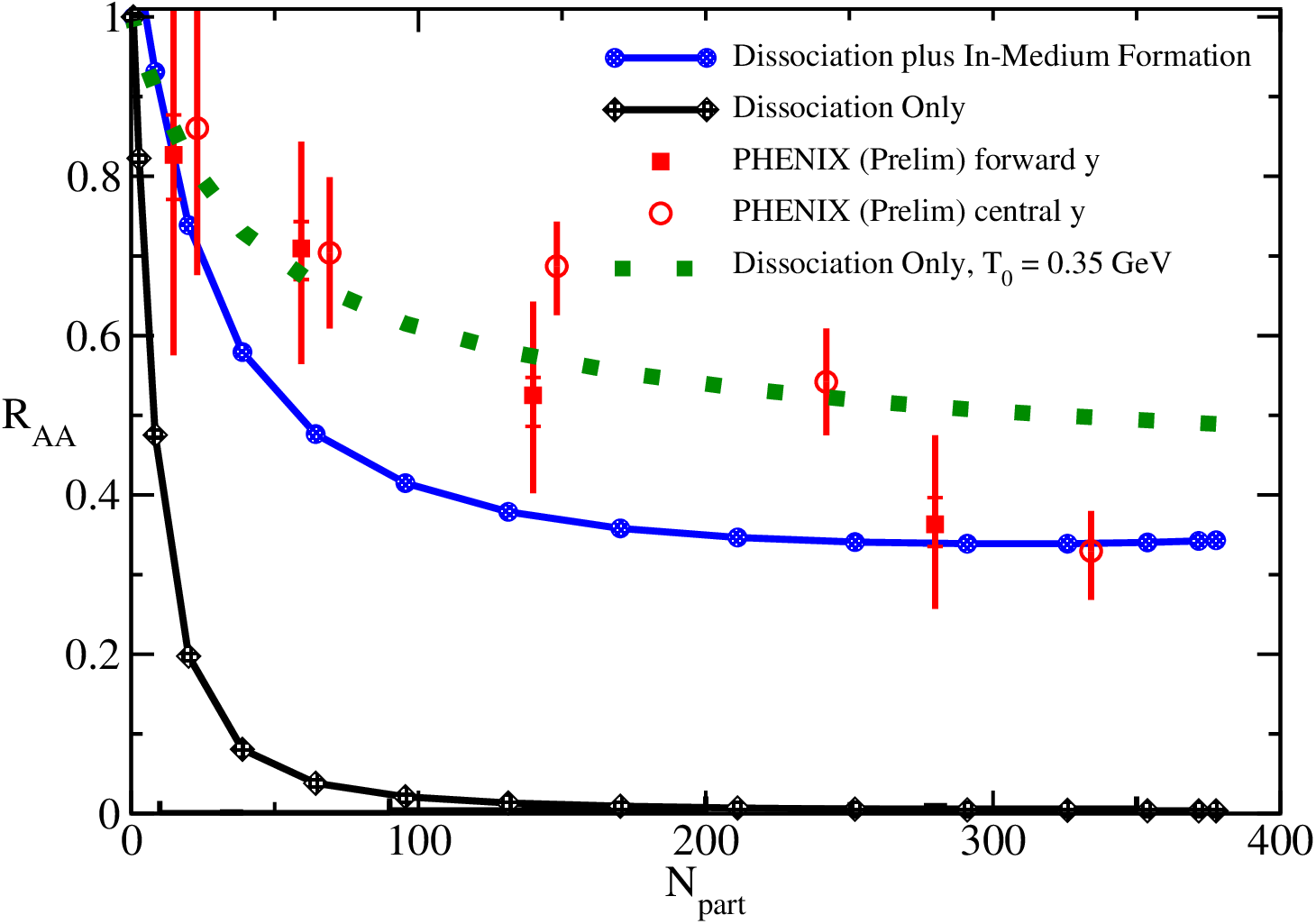}
\caption{\small Centrality dependence of kinetic model without 
in-medium formation.}
\label{noformationRAA}
\end{minipage}
\end{figure}

We show in Fig. \ref{InitialPHENIXdata} the initial PHENIX 
data \cite{Adler:2004zn}, which
consisted of measurements of $d\NJ/dy$ at central rapidity in three 
centrality intervals.  Within limited
statistics, it is evident that there is suppression below even
binary scaling.  However, the flexibility of this model is quite
substantial, as shown in Fig. \ref{parameterspaceconstraint}.
The most recent PHENIX data for $\J$ production in Au-Au interactions
at 200 $\gev$ is presented in terms of \RAA, in which the normalizing factor
is the equivalent binary-scaled $\J$ yield from a superposition of pp 
collisions.  We show in Fig. \ref{kineticRAApredictions} predictions of the
kinetic model using the parameter set as shown in 
Fig \ref{InitialPHENIXdata}.  Our model calculations are limited to
normalized spectra, so that predictions for $R_{AA}$ are uncertain up
to an overall magnitude.  For purposes of comparison with data, we
normalize to unity at the most peripheral $\Npart$.
One sees that the suppression of initially-produced $\J$ is almost
complete, which is
controlled by the initial temperature parameter $T_0$ = 0.5 $\gev$ in
this case.  Inclusion of
the in-medium contribution is seen to be generally consistent
with the data within present uncertainties.  For contrast, we insert 
in Fig. \ref{noformationRAA} the model predictions without in-medium
formation, using a much lower $T_0$ = 0.35 $\gev$.  One sees that
this scenario is also roughly consistent with the data.  Thus we are in
a situation where one needs additional input on an ``expected"
baseline which contains no in-medium formation, in order to make
a clear case for its presence or absence.

Fortunately, the $\J$ momentum spectra are also measured, and they
contain independent signatures with which to compare.

\section{Normalized {\bf $\J$}  y and {\bf \pt} spectra}

We present here a brief overview of the method used to extract these
spectra.  The details are contained in Ref. \cite{Thews:2005vj}.
The starting
point is a version of Eq. \ref{formationintegral} which is 
differential in $\J$ momentum. For the in-medium formation contribution,
one obtains

\begin{equation} 
\frac{dN_{\J}}{d^3 P_{\J}} = \int{\frac{dt}{V(t)}}
\sum_{i=1}^{N_c} \sum_{j=1}^{N_{\bar c}} {\it {v}_{rel}} 
\frac{d \sigma}{d^3 P_{\J}}(P_c + P_{\bar{c}} \rightarrow P_{\J} + X),
\label{formdist}
\end{equation}  where the sum over all $\ccbar$ pairs incorporates
the total formation reactivity.  Since we are only interested in the
normalized spectra, the overall magnitude factors which proved to be
extremely sensitive to model parameters can be disregarded.  We
just calculate the relative formation probability using a model
cross section and evaluate the relative formation rates from 
each pair.  The quark momentum distributions were obtained by
generating a sample of $\ccbar$ pairs directly from NLO pQCD
amplitudes.  Fortunately, we find that the resulting normalized
spectra are quite insensitive to the details of the formation
cross section.  In addition, we have shown \cite{Thews:2005vj} that
final state dissociation effects have a negligible effect on the 
normalized formation spectra.

\begin{figure}[htb]
\begin{minipage}[t]{80mm}
\epsfig{clip=,width=7.8cm,figure=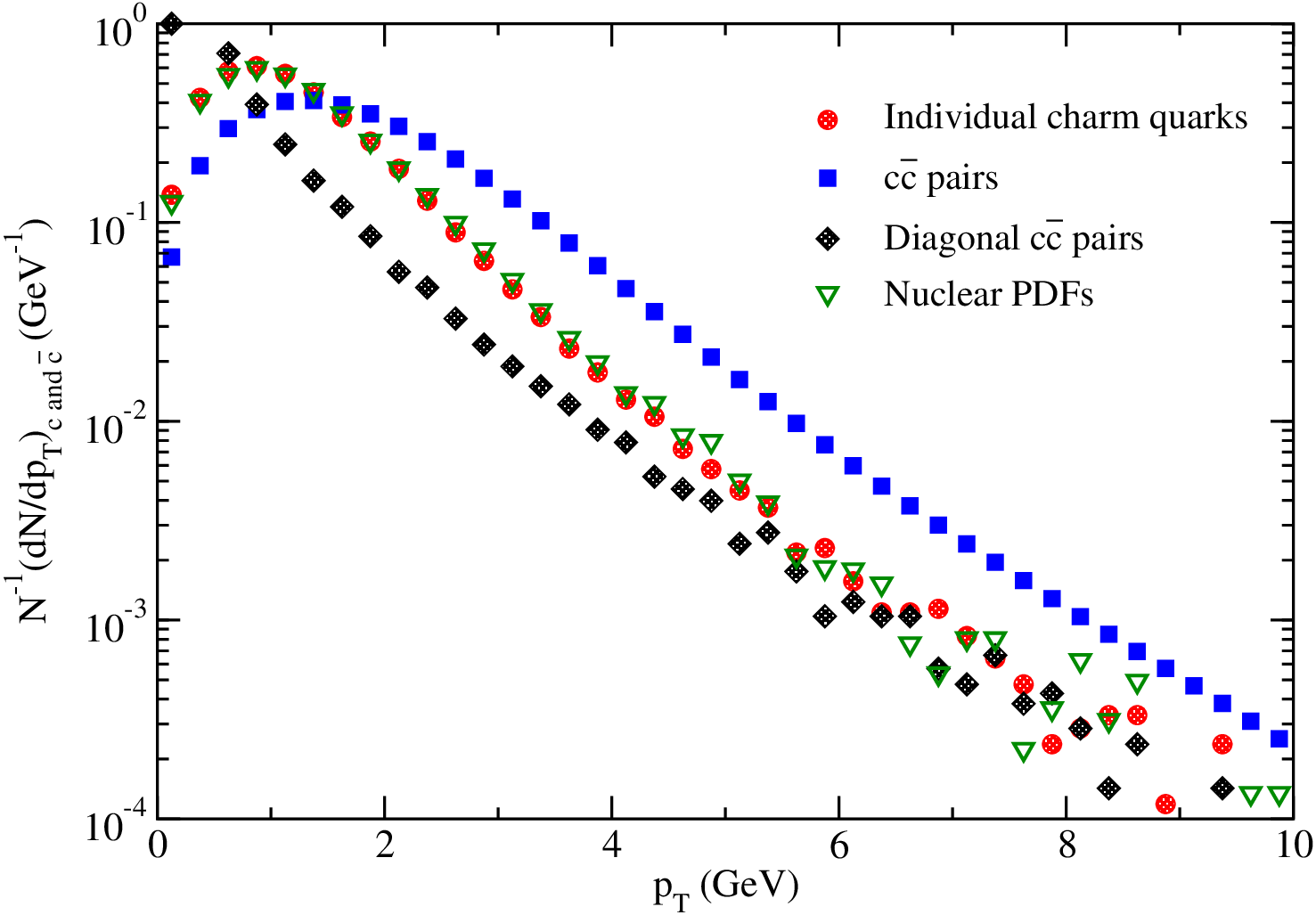}
\caption{\small Charm quark and pairs\\ \pt distributions from
pQCD amplitudes.}
\label{cquarkpt}
\end{minipage}
\hspace{\fill}
\begin{minipage}[t]{80mm}
\epsfig{clip=,width=7.5cm,figure=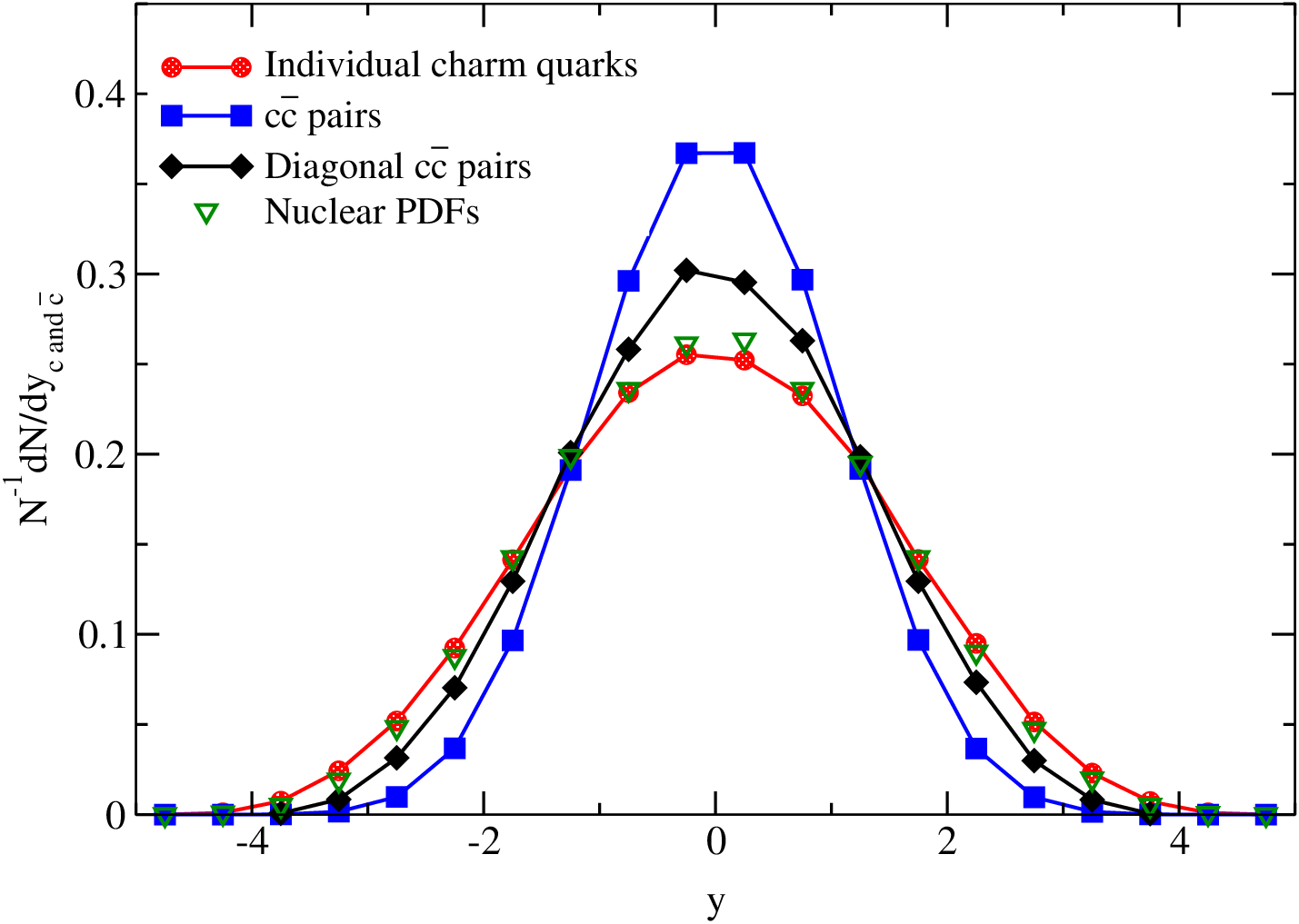}
\caption{\small Charm quark and pairs rapidity distributions from 
pQCD amplitudes.}
\label{cquarky}
\end{minipage}
\end{figure}

The key signature is generated by utilizing
appropriate subsets of the $\ccbar$ pairs.  For no in-medium
formation, we include only the ``diagonal" pairs, i.e. those 
which were generated in the same parton-parton interaction from
pQCD amplitudes.  In this case we employ a color-evaporation type
of scenario, in which the relative probability of hadronization
into a $\J$ is independent of the quark pair momentum.  For the
in-medium formation, we include also the ``off-diagonal" pairs,
in which a charm quark from one interaction forms the $\J$ by
combining with an anticharm quark which was produced in
a different pQCD interaction.  Thus the difference between
$\J$ spectra with and without in-medium formation arises from 
a combination of two effects: the different pair momentum
distributions of the diagonal vs. off-diagonal pair sample, and
the inclusion of a formation probability for the formation
reaction in the medium.  This behavior is shown in 
Fig. \ref{cquarkpt} and Fig. \ref{cquarky}.
The effect of in-medium formation on the rapidity spectra is
quite simple. The dominant factor comes from the weighting
of the $\ccbar$ pairs by the formation probability, which
has previously been shown to predict a narrowing of the 
rapidity distribution.  Initial PHENIX data \cite{PereiraDaCosta:2005xz}
is now available, 
and does not show any evidence for narrowing of the spectra.  
However, the uncertainties are still quite large, so that a
definitive statement cannot be made at this time. For now, we
proceed with a detailed analysis of the transverse momentum spectra.

\section{Initial state effects and centrality dependence of {\bf $\pts$}}

The charm quark spectra we calculate from pQCD have used
collinear parton interactions only.  To simulate the effects of
confinement in nucleons, and more importantly the initial state
\pt broadening due to multiple scattering of nucleons, we have
supplemented the quark pair \pt distributions by adding a 
transverse momentum kick to each quark in a diagonal pair, 
chosen from a Gaussian distribution with width characterized by
$\kts$.  This effect is magnified in the difference between diagonal and
off-diagonal pairs.  Since the azimuthal direction of the $\vec{k}_T$
is uncorrelated from pair-to-pair, an off-diagonal pair with initial
$\pts$ will increase to $\pts +  2 \kts$.  However, the azimuthal correlation
inherent in the diagonal pair will result in an increase to
$\pts +  4 \kts$. This effect will be evident when we consider the
centrality dependence of \pt spectra.

We next use reference data \cite{Adler:2005ph} on $\J$ production in pp and d-Au 
interactions to determine
some of these parameters.  A fit to the dimuon data \pt spectra fixes
$\kts_{pp} = ~0.5 \pm 0.1\; \gev^2$.  We determine the initial state
nuclear \pt broadening with a standard random walk picture.
\begin{equation}
\pts_{pA} - \pts_{pp}\; = \lambda^2\; [\bar{n}_A - 1], 
\label{pApt}
\end{equation}
where $\bar{n}_A$ is the average number of inelastic interactions
of the projectile proton as it
traverses the nucleus A, and $\lambda^2$ is proportional 
to the square of the
transverse momentum transfer per collision. For a nucleus-nucleus collision,
the corresponding relation is
\begin{equation}
\pts_{AB} - \pts_{pp}\; = \lambda^2 \;[\bar{n}_A + \bar{n}_B- 2]. 
\label{AApt}
\end{equation}
We now update the determination of the parameter $\lambda^2$ from that
in Ref. \cite{Thews:2005vj}.  
This update uses a Woods-Saxon nuclear density
parameterization, rather than the sharp sphere model used previously.
We should note that $\bar n$ and $\lambda^2$ are correlated within 
a given nuclear geometry.  We use  the published \cite{Adler:2005ph}
 $\pts_{pp} 
= 2.51 \pm 0.21$ and the average of 
$\pts_{d-Au}= 4.28 \pm 0.31 (y=-1.7)$ and 
$3.63 \pm 0.25 (y = +1.8)$ and
calculate $\bar{n}_A = 3.56$ for minimum bias d-Au interactions at RHIC
energy (using $\sigma_{pp}$ = 42 mb), to extract
$\lambda^2 = 0.56 \pm~0.08~\gev^2$.  
Given this parameter, one can use calculated values of $\bar{n}_{Au-Au}$
as a function of centrality to predict $\pts$ for the $\J$ which
originate from initially-produced diagonal $\ccbar$ pairs.

\begin{figure}[htb]
\epsfig{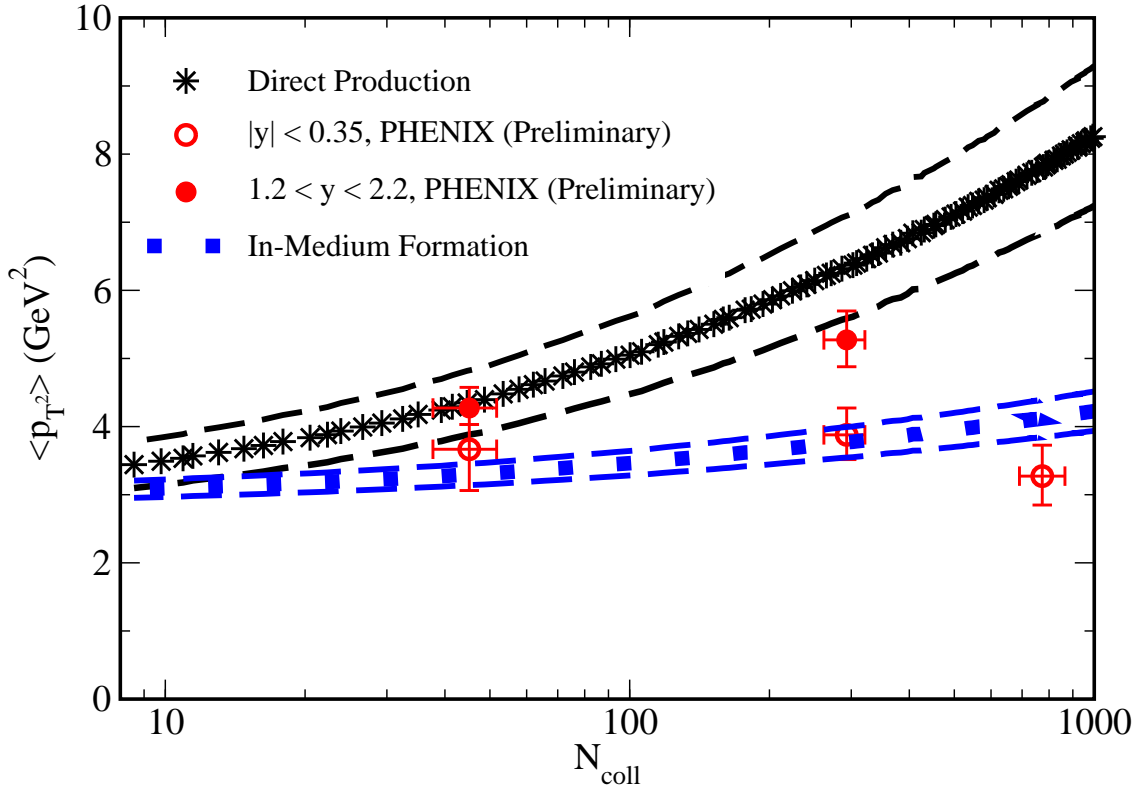}
\caption{\small Centrality dependence of $\pts$ comparing
initial (direct) production and in-medium formation with
preliminary PHENIX data.}
\label{directvsformationpt2}
\end{figure}

An extension of this procedure is then used to predict $\pts$ as a function
of centrality for the $\J$ which originate from in-medium formation. Here there
are two effects which contribute.  One involves the same 
(centrality-dependent) intrinsic $\kts$
due to initial state effects in AA interactions, 
but which now enters with a different
numerical coefficient for off-diagonal pairs.  The other follows the
$\pts$ of the in-medium formation process, which has its own
characteristic dependence on this same
$\kts$. Results of this calculation are shown in Fig. \ref{directvsformationpt2}.
The centrality measure is parameterized by the number of binary collisions, also
calculated in the Glauber formalism using the Woods-Saxon density 
parameterization.  
The stars and dashed bars indicate the central value of the
model predictions for the two scenarios.
One sees clearly that the centrality dependence of $\pts$ is
significantly different for the $\J$ produced initially as 
compared with those which were formed in-medium.  Thus 
measurements in very central collisions will have an enhanced
ability to differentiate between these two mechanisms of
$\J$ production.  The dashed lines indicate the uncertainty in
the model calculations, primarily due to the value of the parameter
$\lambda^2$ determined from the broadening in dAu interactions, 
but also including the uncertainty in the measured $\pts$ in pp
interactions.

\begin{figure}[htb]
\begin{minipage}[t]{80mm}
\epsfig{clip=,width=7.8cm,file=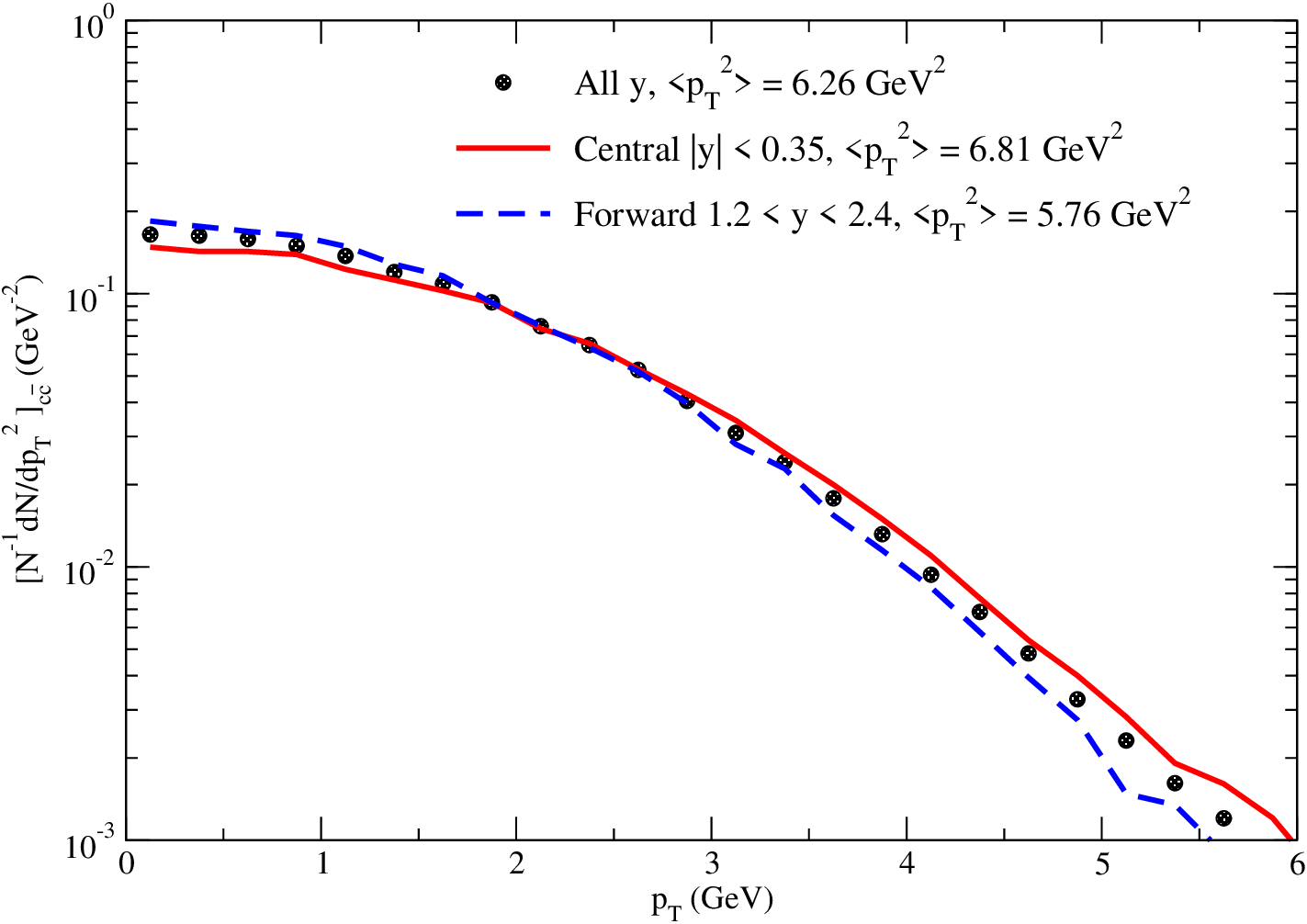}
\caption{\small Rapidity interval variation \\of $\pts$ of diagonal
$\ccbar$ pairs produced in \\Au-Au interactions.}
\label{ybindiagonalccbar}
\end{minipage}
\hspace{\fill}
\begin{minipage}[t]{85mm}
\epsfig{clip=,width=7.8cm,file=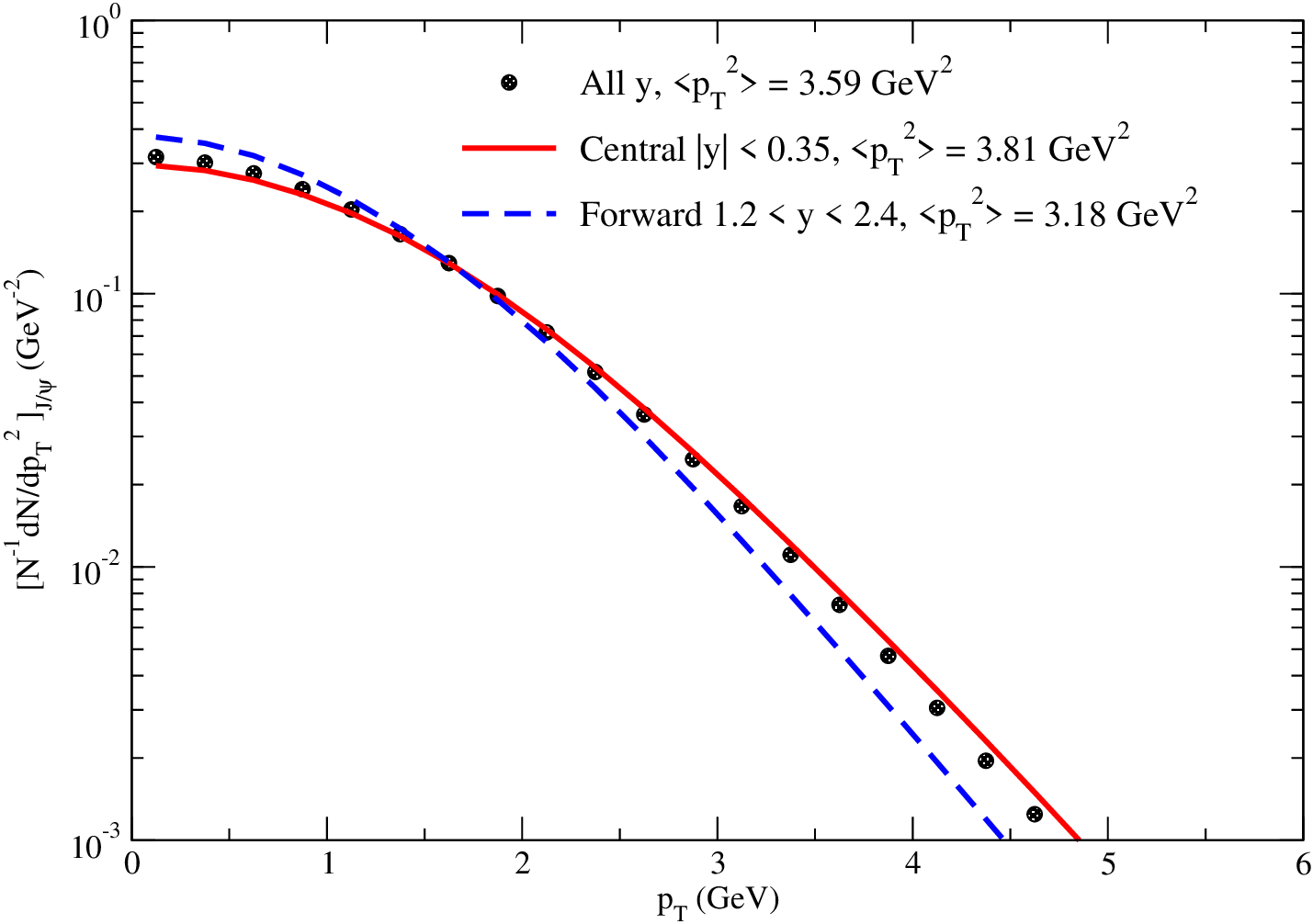}
\caption{\small Rapidity interval variation of $\pts$ for $\J$
which are formed in-medium.}
\label{ybinformation}
\end{minipage}
\end{figure}

The preliminary PHENIX data \cite{PereiraDaCosta:2005xz}
 for $\pts$ is also shown in Fig. \ref{directvsformationpt2}
for three centrality intervals, spanning 0-20\%, 20-40\%, and
40-90\%.  Although the experimental uncertainties make a
definitive conclusion difficult, it appears that the 
in-medium formation predictions may be somewhat more consistent
with the data.  Unfortunately the most central point for the forward
rapidity region is not available, but a corresponding
measurement for backward rapidity may be forthcoming.
Measurements in the central rapidity
interval and the forward rapidity interval appear to differ in
the 20-40\% centrality interval, which
would be difficult to interpret in terms of the initial production
vs. in-medium formation scenario we have presented. However, the
model calculations have
been performed over the entire rapidity region, and thus one 
can look for differences within the experimental rapidity
intervals.  Fig. \ref{ybindiagonalccbar} shows that this variation
for the diagonal $\ccbar$ pairs is much smaller than that for the
measurements, and Fig. \ref{ybinformation} exhibits the same
situation for the in-medium formation scenario.  Thus for a definitive 
conclusion to be reached in this model comparison, one will require
not only a decreased experimental uncertainty in the measured 
$\pts$, but will require that either the dependence on rapidity interval
must essentially disappear, or that some rapidity-dependent final state
effect dominates the spectra.
\section{Summary}
We have shown that the width of $\J$~\pt spectra in AA interactions can
differentiate between initial production and
in-medium formation via pQCD charm quarks.  At large centrality, one
requires only a measurement of $\pts$ with less than approximately
20\% uncertainty.  The present differences between spectra using
measurements at central and forward rapidity are puzzling within this
framework. The corresponding spectra for in-medium formation using
thermal charm quark distributions, as well as direct hadronization of
$\J$ with a thermal distribution \cite{Thews:2005vj} 
result in $\pts$ substantially smaller
than either the preliminary data or the two model results presented
here. It remains to be seen if a combination of such production
processes can be made consistent with the data.   
\ack
This work was partially supported  by a grant from
the U.S. Department of Energy,  DE-FG02-04ER41318.


\section*{References}
\begin{thebibliography}{99}

\bibitem{jpsiform}
Thews R L, Schroedter M and Rafelski J 2000
{\it Phys. Rev.} C {\bf 63} 054905

\bibitem{PereiraDaCosta:2005xz}
  Pereira Da Costa H 2005  (PHENIX Collaboration)
  {\it Preprint} nucl-ex/0510051.

\bibitem{Karsch:2005nk}
  Karsch F, Kharzeev D and Satz H 2006
  {\it Phys. Lett.} B {\bf 637} 75


\bibitem{Thews:2005vj}
  Thews R L and ~Mangano M L 2005
  {\it Phys. Rev.} C {\bf 73} 014904

\bibitem{Adams:2004fc}
  Adams J {\it et al.}  2005 (STAR Collaboration)
  {\it Phys. Rev. Lett.} {\bf 94} 062301 

\bibitem{Adler:2004ta}
  Adler S S {\it et al.} 2005  (PHENIX Collaboration)
  {\it Phys. Rev. Lett.}  {\bf 94} 082301

\bibitem{Adler:2004zn}
  Adler S S {\it et al.}  2005 (PHENIX Collaboration)
  {\it Phys. Rev.} C {\bf 71} 034908, 
  Erratum-ibid. C {\bf 71} 049901 

\bibitem{Adler:2005ph}
  Adler S S \etal 2006 (PHENIX Collaboration) {\it Phys. Rev. Lett}
  {\bf 96} 012304

\end{thebibliography}
\end{document}